\documentclass[aps,twocolumn, pra]{revtex4-1}

\usepackage{newlfont}
\usepackage{color}           
\usepackage{soul}            
\usepackage{amssymb}
\usepackage{amsfonts}
\usepackage{amsmath}
\usepackage{wasysym}
\usepackage{graphicx}
\usepackage{epsfig}
\usepackage{amsthm}
\usepackage{bm}
\usepackage[colorlinks=true, citecolor=blue, urlcolor=blue ]{hyperref}

\def\DJo{$\;$\kern-.4em \hbox{D\kern-.8em\raise.15ex\hbox{--}\kern.35em okovi\'c}}

\begin{document}

\title{Maximally dense coding capable quantum states}

\author{Rabindra Nepal\(^{1}\), R. Prabhu\(^{2}\), Aditi Sen(De)\(^{2}\), and Ujjwal Sen\(^{2}\)}

\affiliation{\(^1\)School of Physical Sciences, Jawaharlal Nehru University, New Delhi 110 067, India\\
\(^2\)Harish-Chandra Research Institute, Chhatnag Road, Jhunsi, Allahabad 211 019, India}

\begin{abstract}
A complementarity relation is established between the capacity of multiport classical information transmission via quantum states and multiparty quantum correlation measures for three-qubit pure states. The multiparty quantum correlation measures considered are the generalized geometric measure, the tangle, and the discord monogamy score. The complementarity relation is revealed by the identification of a one-parameter family of pure three-qubit states, which we call the maximally dense coding capable family of states. These states have the maximal multiport dense coding capacity among all three-qubit pure states with an arbitrary fixed amount of the multiparty quantum correlations.

\end{abstract}  

\maketitle

\section{Introduction}
Quantum correlation \cite{HHHH,DiscordRMP} is known to be a necessary ingredient of most quantum information protocols that are  advantageous over their classical counterparts. Such protocols include quantum communication processes \cite{comm-review} like quantum dense coding \cite{BW}, quantum teleportation \cite{teleportation}, quantum cryptography \cite{Ekert91}, as well as  quantum computational tasks \cite{JozsaEkertRMP} like the one-way quantum computer \cite{Briegel} and deterministic quantum computation with one quantum bit \cite{Knill-Laflamme}. Such theoretical breakthroughs have been closely followed by experimental realizations of these protocols in several physical systems like photons, trapped ions, atoms in optical lattices, nuclear magnetic resonance, etc. \cite{exp}.


These developments necessitate the characterization and quantification of quantum correlations of quantum states shared between separated observers. However, the understanding of entanglement of multisite systems, i.e., systems of more than two particles is still limited, due to its structural complexity \cite{Ashimony,Mbalsone,Dameyer,Osterloh,GGM,VedralPlenio}. This is also true for information-theoretic quantum correlation measures \cite{others,workdeficit, Modi,Arun-disso,discord-multi,bakisob, discord}.
However, it is well known that the commercialization of quantum information protocols requires the realization of multiparticle quantum correlated systems. For example, a quantum computer is more powerful than its classical counterpart, only when quantum coherence can be built among at least thousands of accessible qubits \cite{ciraczollarPreskill}. 

Multipartite quantum correlation can be quantified in a large variety of ways. An important one is by using the concept of monogamy of shared {\em bipartite} quantum correlations in a multiparty quantum system \cite{Ekert91,BenMono,Coffmankunduwotters}. Monogamy is an important connecting theme among the bipartite quantum correlation measures, and in simple terms requires that if two particles share a quantum state with a high quantum correlation, they cannot have a significant amount of quantum correlation with any third particle. Multiparty quantum correlation measures, that are constructed via the concept of monogamy of a bipartite quantum correlation, has the added benefit of providing information about the sharability of bipartite quantum correlations in the corresponding multiparty system. Such multiparty quantum correlations can be formulated by using a variety of bipartite quantum correlation measures, where the latter can belong to the entanglement-separability paradigm \cite{HHHH} or the information-theoretic one \cite{DiscordRMP}. Below we consider two such measures, one from each paradigm, and are respectively called the tangle \cite{Coffmankunduwotters} and the discord monogamy score \cite{amadermono,amaderlightcone, amadervanishingdiscord}. Multipartite quantum correlations can also be quantified independently of the monogamy concept, an example being the generalized geometric measure (GGM), which is  measure of genuine multiparty entanglement in multiparty quantum states. Importantly, the GGM can be efficiently computed for pure states of an arbitrary number of parties in arbitrary dimensions \cite{GGM, GGMrefs, amadervanishingdiscord} (cf. \cite{Ashimony}). 

It is important to establish a connection between the multiparty entanglement content of multipartite quantum states with their ability to act as substrates in quantum information protocols. A paradigmatic example of a quantum information protocol is quantum dense coding \cite{BW,dcgeneral}, where an observer sends classical information to another observer,  by using a shared quantum state. 
The protocol has been generalized to the case of multiple ports \cite{dcamader}. As mentioned above, it is the case of multiple ports that is potentially more useful in applications of the quantum dense coding protocol. In the present article, we show that for a given amount of multiparty quantum correlation in any three-qubit pure state, its ability to act as a channel for multiport classical information transmission via quantum states is bounded above by that of a one-parameter class of three-qubit states, which we call the maximally dense coding capable (MDCC) family of states. More precisely, for any three-qubit pure state, with an arbitrarily fixed amount of multiparty quantum correlation, we show that its quantum advantage in multiport dense coding is always lower than that of the MDCC state with the same amount of the multiparty quantum correlation. We prove the result in the cases when the multiparty measure is the GGM or the tangle or the discord monogamy score, 
The results lead to a generic complementarity relation between the quantum advantage in multiport dense coding and multipartite quantum correlation for three-qubit pure states. 

The paper is organized as follows. In Sec. \ref{esmeasures}, we introduce the multipartite quantum correlation measures, viz., the generalized geometric measure, tangle, and  discord monogamy score. 
In Sec. \ref{qdcdca}, we set the notation for the multiport quantum dense coding protocol, its capacity, and 
the corresponding quantum advantage. In Sec. \ref{esdcadv}, we establish the complementarity relation between the quantum dense coding advantage and the various multiparty quantum correlation measures. 
The conclusions on our results are sketched in Sec. \ref{conc}.


\section{Multipartite quantum correlation measures}
\label{esmeasures}


In this section, we briefly discuss about the multipartite quantum correlation measures that will be required later in the paper. They are the generalized geometric measure, tangle, and discord monogamy score. The first two measures belongs to the entanglement-separability paradigm, while the third one belongs to the information-theoretic one. 
We will also discuss the concept of monogamy of quantum correlations. While the first measure is based on the concept of distance to a relevant class of states, the other two are based on the concept of monogamy.

\subsection{Generalized geometric measure}

A multipartite pure quantum state $|\psi_{A_1,A_2,\ldots, A_N}\rangle$, shared between $N$ parties, $A_1,A_2,\ldots, A_N$, is said to be genuinely multipartite entangled if it is not separable across any  bipartition. To quantify the genuine multipartite entanglement for these states, one can define an entanglement measure based on the distance from the set of all multiparty states that are not genuinely multiparty entangled. In this spirit, the generalized geometric measure \cite{GGM} is defined as 
\[
{\delta _G} (|\psi_{A_1,A_2,\ldots, A_N}\rangle) = 1 - \max_{|\phi\rangle} |\langle \phi | \psi_{A_1,A_2,\ldots, A_N}\rangle|^2,
\]
where the maximization is performed over all states $|\phi\rangle$ which are not genuinely multipartite entangled. It turns out that it is possible to compute the GGM efficiently for quantum states of an arbitrary number of parties in arbitrary dimensions, by using the following result \cite{GGM}. We have 
\begin{eqnarray}
\delta_G(|\psi_{A_1,A_2,\ldots, A_N}\rangle) &=& 1 - \max \{\lambda^2_{{\cal A}: {\cal B}} | {\cal A} \cup  {\cal B} = \nonumber \\
& &  \{A_1,A_2,\ldots, A_N\},\, {\cal A} \cap  {\cal B} = \emptyset\},\,\,\,\,\,\,\,
\end{eqnarray}
where \(\lambda_{{\cal A}:{\cal B}}\) is  the maximal Schmidt coefficient in the \({\cal A}: {\cal B}\) bipartite split  of \(|\phi \rangle\).

\subsection{Tangle}
Unlike the GGM, the multipartite quantum correlation measures in this and the following subsections are based on the concept of monogamy. For a given quantum correlation measure, ${\cal Q}$, and for an arbitrary three-party quantum state $\rho_{ABC}$, shared between Alice $(A)$, Bob $(B)$, and Charu $(C)$, the amount of quantum correlation shared between the Alice-Bob pair and the Alice-Charu pair is typically restricted by the monogamy of quantum correlations \cite{Ekert91,BenMono,Coffmankunduwotters}. The sum of the quantum correlations of the $AB$ and $AC$ pairs is therefore non-trivially bounded. Note here that we have arbitrarily assigned Alice as the ``nodal observer''. A three-party quantum state $\rho_{ABC}$ is said to be monogamous with respect to a quantum correlation measure ${\cal Q}$, if the nontrivial bound is ${\cal Q}_{A:BC}$ \cite{Coffmankunduwotters}, i.e., if it satisfies the inequality 
%
%
%
%
%
\begin{equation}
{\cal Q}_{AB} + {\cal Q}_{AC} \leq {\cal Q}_{A:BC}.
\label{eq:qmonogamy}
\end{equation}
Here ${\cal Q}_{AB}$ is the quantum correlation between the Alice-Bob pair, ${\cal Q}_{AC}$ is the same for the Alice-Charu pair, and ${\cal Q}_{A:BC}$ is the quantum correlation between Alice and the Bob-Charu pair. 
The tripartite physical quantity 
\begin{equation}
\delta_Q = {\cal Q}_{A:BC}-{\cal Q}_{AB}-{\cal Q}_{AC},
\label{eq:qmonogamy}
\end{equation}
known as the quantum monogamy score (corresponding to the measure ${\cal Q}$) \cite{Coffmankunduwotters,amadermono,amaderlightcone, amadervanishingdiscord}, can be expected to quantify the tripartite quantum correlations in the state $\rho_{ABC}$. 

The tangle is the quantum monogamy score corresponding to the square of the bipartite entanglement measure called concurrence, which we define now briefly. The concurrence  quantifies the amount of entanglement present in a two-qubit state and it originated from the  entanglement of formation \cite{BenMono,Wootters}. 
For a two-qubit state $\rho_{AB}$, shared between $A$ and $B$, the concurrence is defined as
$$C(\rho_{AB}) = \max\{0,\lambda_1-\lambda_2-\lambda_3-\lambda_4\},$$
where the $\lambda_i$ are the square roots of the eigenvalues of $\rho_{AB}\tilde{\rho}_{AB}$ in decreasing order. $\tilde{\rho}_{AB}= (\sigma_y \otimes \sigma_y)\rho_{AB}(\sigma_y \otimes \sigma_y)$, with the complex conjugation being taken in the computational basis. $\sigma_y$ is the Pauli spin matrix.


The tangle for the three-qubit state $\rho_{ABC}$ is given by \cite{Coffmankunduwotters}
\begin{equation}
\delta_{\tau}(\rho_{ABC}) = C^2_{A:BC} - C^2_{AB} - C^2_{AC}.
\label{eq:tangle}
\end{equation}

For notational convenience, we have denoted the tangle as $\delta_{\tau}$. Note that we have defined the tangle for three-qubit states only.  The tangle is monogamous for all three-qubit states \cite{Coffmankunduwotters,monogamybunch}.

\subsection{Discord monogamy score}

In the preceding subsections, we have defined multiparty quantum correlation measures that fall within the entanglement-separability paradigm. We now define a multiparty quantum correlation measure based on information-theoretic concepts. However, just like for the tangle, we again use the concept of monogamy. To proceed further, we have to define the bipartite information-theoretic quantum correlation measure called quantum discord. Quantum discord is defined, for a two-party quantum state $\rho_{AB}$, as the difference between
%
%
two quantum information-theoretic quantities \cite{discord}: 
\begin{equation}
\label{eq:discord}
Q(\rho_{AB})= {\cal I}(\rho_{AB}) - {\cal J}(\rho_{AB}).
\end{equation}
Classically, both ${\cal I}(\rho_{AB})$  and ${\cal J}(\rho_{AB})$ represent the mutual information and they are equivalent.
Quantum mechanically,  \({\cal I}(\rho_{AB})\), of a bipartite state \(\rho_{AB}\), corresponds to the total correlation, and is given by \cite{qmi}
 (see also \cite{Cerf, GROIS})
\begin{equation}
\label{qmi}
\mathcal{I}(\rho_{AB})= S(\rho_A)+ S(\rho_B)- S(\rho_{AB}),
\end{equation}
where $S(\varrho)= - \mbox{tr} (\varrho \log_2 \varrho)$ is the von Neumann entropy of the quantum state \(\varrho\), and 
 \(\rho_A\) and \(\rho_B\) are the reduced density matrices of  \(\rho_{AB}\).
On the contrary, \({\cal J}(\rho_{AB})\) is argued to be a measure of classical correlations \cite{discord} and 
is defined as 
\begin{equation}
\label{eq:classical}
 {\cal J}(\rho_{AB}) = S(\rho_A) - S(\rho_{A|B}). 
\end{equation}
The conditional entropy, \(S(\rho_{A|B})\), is obtained by performing a projection-valued measurement on \(B\) and is given by
\begin{equation}
S(\rho_{A|B}) = \min_{\{B_i\}} \sum_i p_i S(\rho_{A|i}).
\end{equation}
The measurement is performed on the complete set of rank-one projectors $\{B_i\}$ and the probability of obtaining the outcome $B_i$ is given by 
\(p_i = \mbox{tr}_{AB}[(\mathbb{I}_A \otimes B_i) \rho_{AB} (\mathbb{I}_A \otimes B_i)]\), with
\(\mathbb{I}\) being the identity operator on the Hilbert space of \(A\).
The output state at $A$ corresponding to the outcome \(\{B_i\}\) is 
\(\rho_{A|i} = \frac{1}{p_i} \mbox{tr}_B[(\mathbb{I}_A \otimes B_i) \rho_{AB} (\mathbb{I}_A \otimes B_i)]\).

The discord monogamy score was introduced as a multipartite quantum correlation measure by using the monogamy considerations of the quantum discord in Refs. \cite{amadermono,amaderlightcone, amadervanishingdiscord}. 
It is defined as the quantum monogamy score corresponding to quantum discord:
\begin{equation}
\delta_D = D_{A:BC} - D_{AB} - D_{AC}.
\label{eq:}
\end{equation}
Note that unlike the square of the concurrence, quantum discord for three-qubit states can be both monogamous and non-monogamous \cite{amadermono,amaderlightcone, amadervanishingdiscord}.


\section{Quantum dense coding capacity and the quantum advantage}
\label{qdcdca}

Quantum dense coding is a quantum communication protocol by which an observer can send classical information, beyond a certain ``classical limit'', by using shared entanglement \cite{BW}.
Suppose a sender, Alice, and a receiver, Bob, share an arbitrary quantum state $\rho_{AB}$, defined in the Hilbert space ${\cal H}_A \otimes {\cal H}_B$. The dimension of ${\cal H}_A$ and ${\cal H}_B$ are respectively $d_A$ and $d_B$. It was shown \cite{BW, dcgeneral, dcamader} that the 
amount of classical information that can be sent by Alice to Bob is given by 
\begin{eqnarray}
 {\cal C} (\rho_{AB}) &=& \mbox{max}\{\log_2 d_A,\, \log_2 d_A + S(\rho_B) - S(\rho_{AB})\},\,\,\,\,\,\,\,\,\,\,\,\,
\end{eqnarray}
where the resources necessary are the shared quantum state $\rho_{AB}$ and a noiseless quantum channel for $d_A$-dimensional quantum states from Alice to Bob. Without the use of the shared quantum state, Alice will be able to send only $\log_2 d_A$ bits of classical information. Using the shared quantum state is therefore advantageous only if the ``coherent information'', i.e. $S(\rho_{B})-S(\rho_{AB})$, is positive. The corresponding quantum state is then referred to as ``dense codeable''. $C(\rho_{AB})$ is the capacity of quantum dense coding.

Consider now a multiport situation with a single sender, Alice $(A)$, and $N$ receivers, called Bobs and denoted as $B_1,B_2,\ldots,B_N$, and suppose that they share a multipartite quantum state $\rho_{AB_1B_2\ldots B_N}$. We consider a situation where $A$ wants to send classical messages to all the Bobs individually by using the multiparty shared state $\rho_{AB_1B_2\ldots B_N}$ in a dense coding protocol.
%
%
%
%
The quantum advantage of such a multiport dense coding protocol can be defined as
\begin{equation}
 {\cal C}_{adv} = \mbox{max}[\{S_{B_i} - S_{AB_i}|i = 1, 2, \ldots N\}, 0 ],
\end{equation}
where $S_{B_i}=S(\rho_{B_i})$ and $S_{AB_i}=S(\rho_{AB_i})$. $\rho_{B_i}$ and $\rho_{AB_i}$ are the local density matrices of the state $\rho_{AB_1B_2\ldots B_N}$ for the corresponding parties.

\section{Multipartite quantum correlation measures versus dense coding advantage}
\label{esdcadv}

In this section, we establish a relation between the quantum advantage in the multiport dense coding protocol and multipartite quantum correlation measures for three-qubit pure states. 
We introduce a family of three-qubit pure states, which we call the maximally dense coding capable states. For a given amount of multiparty quantum correlation, the three-qubit pure state that is best suited for multiport dense coding is always from this MDCC family. We begin in subsection \ref{sec:ggmvsdca} with the case when the multiparty quantum correlation is quantified by the GGM. The two subsequent subsections deal with the cases when the measure is respectively the tangle and the discord monogamy score.

\subsection{Generalized geometric measure versus dense coding advantage}
\label{sec:ggmvsdca}

Let us consider three parties, Alice, Bob, and Charu. In this tripartite scenario, we will analytically derive a complementarity relation between the GGM and quantum advantage in multiport dense coding for all three-qubit pure states. Moreover, we will show that the non-trivial boundary of this inequality is traced by the following single-parameter family of three-qubit states:
\begin{equation}
 \label{boundarystate}
|\psi_{\alpha}\rangle = |111\rangle + |000\rangle + \alpha (|101\rangle + |010\rangle), 
\end{equation}
where the state parameter $\alpha$ is assumed to be real. Note that the state displayed in Eq. (\ref{boundarystate}) is not normalized to unity. We call this one-parameter family of quantum states as the MDCC family. Three-qubit pure states are known to be the union of two disjoint classes, viz. the GHZ-class and the W-class, which cannot be reached from each other by stochastic local quantum operations and classical communication \cite{VidalCirac} (cf. \cite{GHZ,W}). They can be distinguished by the values of their tangles, with the GHZ-class states having a (non-zero) positive tangle and the W-class states having a zero tangle. 
%
We find that the tangle of the above state is always positive except at $\alpha =1$ (see Fig. 1). Hence the state
 $|\psi_{\alpha}\rangle$ belongs to the GHZ-class except the point $\alpha =1$ which belongs to the W-class. We now prove the complementarity relation and give the corresponding boundary.

\begin{figure}%
 \includegraphics[width=0.7\columnwidth,height=0.2\textheight]{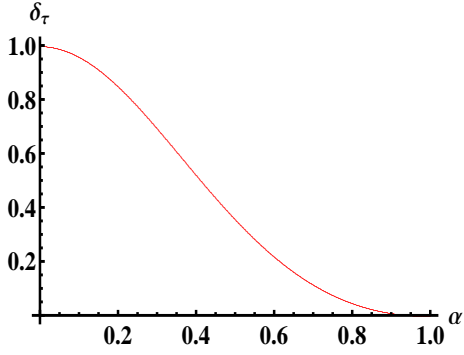}%
 \caption{(Color online.) Tangle of the MDCC family of states. The horizontal axis represents $\alpha$, which is dimensionless, while the vertical axis represents tangle ($\delta_{\tau}$) which is measured in ebits. 
 }
 \label{fig:tangle}%
 \end{figure}

\noindent{\bf Theorem:} \emph{Given an arbitrary three-qubit pure state $|\psi\rangle$ whose GGM is equal to the GGM of $|\psi_{\alpha}\rangle $, the dense coding advantages of $|\psi\rangle $ and $|\psi_{\alpha}\rangle $ always follow the ordering given by}
\begin{equation}
 {\cal C}_{adv} (|\psi_{\alpha}\rangle) \geq {\cal C}_{adv} (|\psi\rangle). 
\end{equation}

\noindent {\bf Proof:} 
The GGM of an arbitrary three-qubit state $|\psi\rangle$ is $ {\delta_G}(|\psi\rangle ) = 1 - \max[\lambda_A, \lambda_B, \lambda_C]$, where $\lambda_A$, $\lambda_B$, and $\lambda_C$ 
are the maximum eigenvalues of the local density matrices for $A$,  $B$, and $C$ respectively of $|\psi\rangle$. On the other hand, the GGM of 
$|\psi_{\alpha}\rangle $ is 
\[
{\delta_G}(|\psi_{\alpha}\rangle ) = \frac{1}{2}- \frac{\alpha}{1 +\alpha^2},
\]
as the local density matrices for $A$ and $C$ of $|\psi_{\alpha}\rangle $ are maximally mixed, while the eigenvalues of that for $B$ of $|\psi_{\alpha}\rangle $ are $\displaystyle{\frac{1\pm 2\alpha}{2(1 +\alpha^2)}}$.

Now, ${\delta_G}(|\psi_{\alpha}\rangle ) =  {\delta_G}(|\psi\rangle )$ implies that 
\begin{equation}
\frac{\alpha}{1 + \alpha^2} = \lambda - \frac{1}{2},
\label{equality}
\end{equation}
 where we call $\lambda$ as maximum of  $\lambda_A$, $\lambda_B$, and $\lambda_C$. 

The advantage of dense coding for $|\psi_{\alpha}\rangle$
is given by 
\begin{equation}
{\cal C}_{adv} (|\psi_{\alpha}\rangle) =S_C-S_B.
\label{eq:advdc}
\end{equation}
For the purpose of this proof $S^{\alpha}_A,\, S^{\alpha}_B,\, S^{\alpha}_C$ denote the von Neumann entropies of the local density matrices for $A,\, B,\,  C$ of $|\psi_{\alpha}\rangle$ and $S_A,\, S_B,\, S_C$ denote those of $|\psi\rangle$. To see Eq. (\ref{eq:advdc}), note that 
%
%
%
\begin{eqnarray}
S^{\alpha}_B = \mbox{tr}_{AC}|\psi_{\alpha}\rangle\langle\psi_{\alpha}|=\mbox{tr}_{B}|\psi_{\alpha}\rangle\langle\psi_{\alpha}|,
\label{eq:}
\end{eqnarray}
and
$$S^{\alpha}_C = \mbox{tr}_{AB}|\psi_{\alpha}\rangle\langle\psi_{\alpha}|=\mbox{tr}_{C}|\psi_{\alpha}\rangle\langle\psi_{\alpha}|,
$$
and that $S^{\alpha}_C-S^{\alpha}_B \geq 0$, as the local density matrix for $C$ of $|\psi_{\alpha}\rangle$ is maximally mixed. We have
\begin{eqnarray}
{\cal C}_{adv} (|\psi_{\alpha}\rangle) &=& 
1 + \left(\frac{1-2\alpha}{2(1 +\alpha^2)}\right) \log_2 \left(\frac{1-2\alpha}{2(1 +\alpha^2)}\right)\nonumber \\
& &+\left(\frac{1+2\alpha}{2(1 +\alpha^2)}\right) \log_2  \left(\frac{1+2\alpha}{2(1 +\alpha^2)}\right).
\label{eq:main}
\end{eqnarray}
The premise, of the equality of the GGMs of $|\psi_{\alpha}\rangle$ and $|\psi\rangle$, implies that 
\begin{equation}
{\cal C}_{adv} (|\psi_{\alpha}\rangle) = 1 - H(\lambda),
\label{eq:candh}
\end{equation}
where $H(p)=-p\log_2 p-(1-p)\log_2 (1-p)$ is the Shannon entropy for the binary probability distribution $\{p,\, 1-p\}$.

Let us now consider the case when $\lambda_B$ is the maximum among all the eigenvalues of $|\psi\rangle$, i.e. $\lambda_B=\mbox{max}\{\lambda_A, \lambda_B, \lambda_C\}$. Hence $\lambda_B \geq \lambda_C$. This, coupled with the fact that $\lambda_B,\, \lambda_C \geq \frac{1}{2}$, immediately implies that $S_C -S_B \geq 0$, and hence $ {\cal C}_{adv} (|\psi\rangle) = S_C - S_B$. Since $S_C \leq 1$, Eq. (\ref{eq:candh})
gives us
\begin{equation}
{\cal C}_{adv} (|\psi_{\alpha}\rangle) = 1 - H(\lambda_B^2) \geq S_C - S_B = {\cal C}_{adv} (|\psi\rangle).
\end{equation}
The derivation is very similar when the maximum eigenvalue is obtained from the local density matrix for $C$ of  $|\psi\rangle$. 

Let us now consider the case when $\lambda = \lambda_A$. Without loss of generality, one can assume that $\lambda_A  \geq \lambda_B \geq \lambda_C$. Moreover $\lambda_A,\, \lambda_B$ and $\lambda_C \geq \frac{1}{2}$. Hence, $S_C - S_B \geq 0$ and
$S_A \leq S_B$. Also $S_C \leq 1$. Therefore, we have 
\begin{equation}
{\cal C}_{adv} (|\psi_{\alpha}\rangle) = 1 - H(\lambda_A) = 1 - S_A \geq S_C - S_B = {\cal C}_{adv} (|\psi\rangle).
\end{equation}
Hence the proof. \hfill $\blacksquare$

The theorem immediately implies the following complementarity relation.

\begin{figure}%
 \includegraphics[width=0.7\columnwidth,height=0.2\textheight]{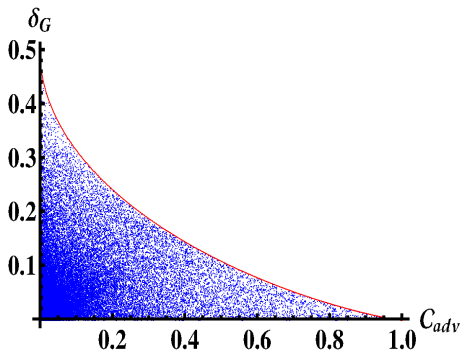}%
 \caption{(Color online.) GGM $(\delta_G)$  vs. dense coding advantage $({\cal C}_{adv})$. The GGM of randomly generated three-qubit pure states is plotted against their quantum advantage of the multiport dense coding. The MDCC family of states represents the boundary of region containing the plotted points. The GGM is plotted on the vertical axis and is dimensionless. The horizontal axis represents the quantum advantage, and is measured in bits.
 }
 \label{fig:DCvsGGM}%
 \end{figure}

\noindent{\bf Corollary:} {\em For three-qubit pure states, the quantum advantage in multiport dense coding and the generalized geometric measure are contained by the complementarity relation}
\begin{equation}
{\cal C}_{adv}+H(\delta_G) \leq 1.
\label{eq:comple}
\end{equation}
To visualize the complementarity relation, we have randomly generated $10^5$ arbitrary three-qubit pure states, and calculated their GGMs and dense coding advantages. A scatter diagram for the corresponding data is given in Fig. \ref{fig:DCvsGGM}. The complementarity relation is clearly visible, with the boundary being given by the equality in Eq. (\ref{eq:comple}). 

%
%

\subsection{Tangle versus dense coding advantage}

The complementarity obtained in the previous subsection has the potential to be generic for an arbitrary multiparty quantum correlation measure. To investigate in this direction, we try to find a similar complementarity between the tangle as a measure of multiparty quantum correlation and the dense coding advantage. For a randomly generated $1.5\times 10^5$ three-qubit pure states, we have plotted a scatter diagram with ${\cal C}_{adv}$ as the abscissae and $\delta_{\tau}$ as the ordinates. See Fig. 3. Quite interestingly, a complementarity relation, which is qualitatively similar to the one obtained before, shows up. Moreover, the boundary is again obtained for the MDCC family of three-qubit states.
Therefore, the numerical simulations imply that the quantum advantage in multiport dense coding, ${\cal C}_{adv}$, and the tangle, $\delta_{\tau}$, are constrained by the complementarity relation
\begin{equation}
{\cal C}_{adv}+H \left(\frac{1}{2}-\frac{1}{2}\sqrt{1-\tau_{\alpha}}\right) \leq 1.
\label{eq:}
\end{equation}



\begin{figure}%
 \includegraphics[width=0.7\columnwidth,height=0.2\textheight]{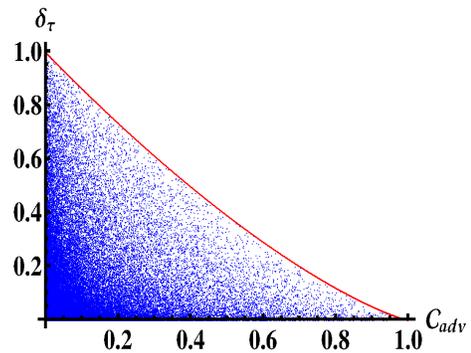}%
 \caption{(Color online.) Tangle $(\delta_{\tau})$  vs. dense coding advantage $({\cal C}_{adv})$. The tangle of randomly generated  $1.5 \times 10^5$ three-qubit pure states is plotted against $({\cal C}_{adv})$.
Again, the boundary of the region containing the plotted points  corresponds to the MDCC family of states. The horizontal axis is the same as in Fig. \ref{fig:DCvsGGM}. The vertical axis represents the tangle, and is measured in ebits. 
 }
 \label{fig:Dcvstangle}%
 \end{figure}


\subsection{Discord monogamy score versus dense coding advantage}
\label{itdcadv}

\begin{figure}%
\includegraphics[width=6cm]{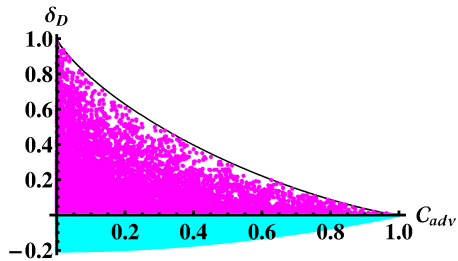}%
\caption{Discord monogamy score $(\delta_D)$ vs. advantage in dense coding protocol $({\cal C}_{adv})$. 
We have randomly generated $6 \times 10^4$ three-qubit states and find that in the upper half in which only the GHZ-class state belongs, the boundary is still represented by $|\psi_\alpha\rangle$. For the region of dense coding advantage and discord monogamy score, the boundary of the W-class states which are always non-monogamous with respect to discord.
}%
\label{DCadvDiscWGHZ}%
\end{figure}

We now investigate whether a similar complementarity relation also holds between ${\cal C}_{adv}$ and discord monogamy score. As has been shown recently, the discord monogamy score, although defined via the concept of monogamy, behaves rather differently than the tangle. In particular, the states from the W-class are always non-monogamous with respect to quantum discord, and so $\delta_D$ is negative for these states \cite{amadermono,amaderlightcone, amadervanishingdiscord}. We have numerically generated $3 \times 10^4$ states from the GHZ-class and a same number of states from the W-class and have plotted their ${\cal C}_{adv}$ and  $\delta_D$ as abscissae and ordinates in a scatter diagram (see Fig. \ref{DCadvDiscWGHZ}). A complementarity relation qualitatively similar to the ones before, shows up once again. And the same MDCC family gives the envelope of the scatter diagram. An additional feature in this case is the spilling over of the plotted points below the ${\cal C}_{adv}$ axis. This is a result of the fact that quantum discord can be non-monogamous for three-qubit states.


\section{Conclusions}
\label{conc}

Quantum dense coding is a quantum communication protocol for sending classical information from one location to another by using a shared  quantum state as a channel. 
Here we have established that in a tripartite scenario, the quantum advantage in multiport dense coding  has a complementary relation with multiparty quantum correlation measures. We have analytically demonstrated that a single-parameter family of tripartite states, that we have called the maximally dense coding capable family of states, is best suited for sending classical information in a multiport dense coding protocol, from among all three-qubit pure states with a given amount of a genuine multiparty entanglement called the generalized geometric measure. Numerical simulations show that the same family of states represents the boundary when the generalized geometric measure is replaced by other multisite quantum correlation measures like the tangle and the discord monogamy score. The broad qualitative features of the complementarity relation remains the same irrespective of the multiparty quantum correlation measure employed.

%
%
%
%

\section*{Acknowledgments}

R.P. acknowledges support from the Department of Science and Technology, Government of India, in the form of an INSPIRE faculty scheme at the Harish-Chandra Research Institute, India. We acknowledge computations performed at the cluster computing facility in HRI.

\end{document}